\documentstyle[12pt,aasms]{article}
\def\etal{et al.\ }
\def\como#1{in BL Lac Objects, ed. L. Maraschi, T. Maccacaro,
\& M.-H. Ulrich (Berlin: Springer-Verlag),
{#1}}
\def\pmax{$P_{\rm m}$}
\begin{document}
\title{The Optical Polarization Properties of X-ray Selected
BL~Lacertae Objects}

\author{Buell T. Jannuzi}
\affil{Institute for Advanced Study, Princeton, NJ 08540; and
Steward Observatory, University of Arizona, Tucson, AZ 85721}

\author{Paul S. Smith}
\affil{Steward Observatory, University of Arizona, Tucson, AZ 85721}

\author{Richard Elston}
\affil{National Optical Astronomy Observatories, CTIO, Casilla 603, La
Serena, Chile 1353}

\begin{abstract}
We discuss the optical polarization properties of X-ray selected BL
Lacertae objects (XSBLs) as determined from three years of monitoring
the polarization of 37 BL Lac objects and candidates.  The observed
objects include a complete X-ray flux limited sample drawn from the
Einstein Extended Medium Sensitivity Survey (EMSS).  We find that the
majority of the XSBLs classified solely on the appearance of the their
optical spectra are true members of the class of BL~Lacertae objects
since they possess intrinsically polarized and variable continua. The
duty cycle of polarized emission (fraction of time spent with the
degree of polarization greater than 4\%) from XSBLs is 40\%.  While
XSBLs have variable polarized emission, the majority ($\approx 85$\%)
have stable or preferred polarization position angles on time scales
at least as long as three years. This reflects stability in the
geometry of the region emitting the linearly polarized optical
emission.  We describe the observed spectral dependence of the degree
of polarization and discuss some of the possible mechanisms producing
the observed characteristics.  While dilution of the polarized
emission by the host galaxy starlight is certainly present in some
objects, we demonstrate that the average polarization properties of
XSBLs derived from our observations are not drastically
affected by this effect. While the confirmed BL Lac objects are shown
to be photometric variables, the objects in our monitored sample did
not display the larger than one magnitude variations generally used to
characterize the optical variability of radio selected BL Lacertae
objects or blazars in general.

\end{abstract}


\section{Introduction}

In 1987 we began a program to measure and monitor the optical
polarization of X-ray selected BL Lacertae objects (XSBLs).  The
majority of \hbox{XSBLs} have received their classification on the
basis of essentially one criterion: that optical spectroscopy of
objects within the X-ray position error box of an X-ray source
revealed an object that presented some evidence of having a
non-thermal continuum by exhibiting a ``featureless'' spectrum.  We
are interested in determining whether these X-ray selected objects are
really members of the class of BL Lac objects, possessing
intrinsically variable polarized emission indicative of optical
synchrotron radiation.  Evidence has been presented indicating that
XSBLs have significant differences in their observed properties from
radio selected BL Lac objects (RSBLs) (Stocke \etal 1985). These
differences included apparently less variable and less polarized
optical emission.  Although the studied sample of XSBLs consisted of
only eight objects, the apparent differences led to questions about
the relationship between the objects selected from surveys made at the
two different bands.

Work by other authors examines the optical spectra, flux variability,
and radio properties of XSBLs (e.g. Stocke \etal 1985; Stocke \etal
1990; Morris \etal 1991; Stocke \etal 1991; Perlman \& Stocke 1993;
Laurent-Muehleisen \etal 1993).  A previous examination of the
polarization properties of an incomplete sample of XSBLs was presented
by Schwartz \etal (1989).  Our work (Jannuzi 1990; presented here and
in two related papers) is the most extensive and systematic
examination of the optical polarization properties of a complete
sample of XSBLs.  Together, these studies provide the data necessary
to compare the radio and X-ray selected objects.

We have divided the presentation of our results into three papers.  In
Jannuzi, Smith, \& Elston (1993, paper I) we present our definitions
of the terms X-ray and radio selected BL Lac, highly polarized quasar
(HPQ), and blazar; describe the observing techniques and procedures
used in obtaining our polarimetry and photometry data; describe the
composition of the X-ray selected and radio selected samples of BL Lac
objects that we will use in our general study of the optical
polarization properties of all BL Lac objects; and present all of our
polarimetry and photometry data.  Jannuzi, Elston, \& Smith (1994,
paper~III) compare the optical polarization properties of XSBLs with
RSBLs as part of our efforts to learn more about the entire class.  In
this paper (paper II), we describe the general optical polarization
properties of XSBLs as derived from our observations.  In \S~2 we
present the general polarization characteristics of XSBLs.  We
describe the detection, maximum observed polarization, variability,
and frequency dependence of the polarization.  We examine the possible
effects of the host galaxy starlight on the observed polarization
properties of BL Lac objects in \S~2.4.  In \S~3 we present the
results of our photometry including a discussion of detected
variability.  For a discussion of the properties of individual objects
see paper I.  We discuss the proper classification of these objects in
\S~4. In \S~5 we summarize the results of this paper.

\section{Polarimetry of XSBLs}

We have measured the polarization of all of the XSBLs in our sample
and monitored their polarization.  Our monitored objects include the
complete X-ray flux limited sample compiled as part of the Einstein
Extended Medium Sensitivity Survey (EMSS, Gioia \etal 1990; Stocke
\etal 1989; Stocke \etal 1990; Stocke \etal 1991; Morris \etal 1991).
The program objects are listed in Table~1 and their selection is
discussed in paper I.  Positions, finding charts, and photometrically
calibrated comparison stars for the monitored objects are presented in
Smith, Jannuzi, \& Elston (1991).  While every effort was made to
observe all of the objects as many times as possible, there are
disparities in the degree objects were monitored.  This was the result
of several constraints.  First, the bulk of our monitoring was
performed with the Steward Observatory (SO) 1.54~m telescope, which
cannot observe objects with declinations greater than $61^\circ$.
Second, the fainter objects (m$_{\rm v} > 19.0$) could be observed at
the SO 1.52~m and 1.54~m telescope only if the weather and seeing were
excellent. Third, objects with low declinations were obviously not as
easy to monitor from Steward observatory as objects at higher
declinations.  Despite these restrictions, we were able to monitor the
majority of the EMSS XSBLs.  We note that our monitoring data are not
biased by the past behavior of the objects in the sample.  It is well
known that the early polarimetry of RSBLs (especially for the
published data) is biased by the ``hot object'' effect.  Objects which
were observed to be highly polarized tended to be more extensively
monitored than objects which had low polarizations when first
observed.  This problem is still affecting our knowledge of the
polarization properties of quasars. In our study of XSBLs we have
attempted to observe all of the observable program objects during each
night of observation, regardless of their past polarization history.
When possible, we have performed multicolor polarimetry in order to
examine the frequency dependence of the percent polarization and
position angle.

\subsection{Are XSBLs Polarized?}

The results of our polarization and photometry observations of the
individual objects are summarized in Table~1\thinspace.  In column (2)
we indicate to which X-ray selected sample an object belongs:
HEAO-A2~=~{\it HEAO-1} A-2 high latitude sample, HEAO-NRL~=~NRL {\it
HEAO-1} catalogue as identified by the Large Area Sky Survey,
EMSS~=~EMSS BL Lac object or candidate BL Lac object, C-EMSS
identifies objects which are members of the complete subsample of EMSS
BL Lac objects (see paper~I for a discussion of these samples).

Prior to our observations, the majority of our program objects had not
been observed for polarized emission.  We observed 31 of these objects
to be polarized on at least one occasion.  We were able to make
significant observations for 21 of the 22 members of the C-EMSS.  We
deem an observation ``significant'' if it yields a three sigma
detection of polarized emission ($P/\sigma_{p} > 3$) or a two sigma
limit of 4 \%.  We define our two sigma limits to be the measured
percent polarization plus two times the uncertainty in the measurement
($P_{\rm obs} + [2\times \sigma_p] < 4$\%, where $P_{\rm obs}$ is the
observed percent polarization without correction for statistical
bias).  We chose a limit of 4\% because we were generally able to
obtain limits of this quality given our observational constraints.

If ``ISP'' appears in column (12), significant interstellar
polarization was detected in the field of the object.  Polarization
measurements were made of stars in the fields of all the observed
XSBLs in order to check for interstellar polarization caused by dust
along the line of sight. The measured stars included the photometry
comparison stars described in Smith \etal (1991).  The results of
these measurements are discussed further in paper~I.  For two objects
(MS~0205.7+3509 and MS~0419.3+1943) the comparison stars were observed
to be polarized with position angles agreeing with the polarization
position angles of the BL Lac object candidates and conclusive
variability in the polarization of these objects was not detected.  We
do not consider either of these objects to have shown intrinsic
polarization.

We also note that the following objects were detected to be polarized
on only one occasion or were only marginal detections: MS~0607.9+7108,
H$\,$1101$-$232, MS$\,$1229.2+6430, MS$\,$1235.4+6315,
MS$\,$2336.5+0517, and MS$\,$2347.4+1924.

\subsection{Variability of Polarization}

Our polarization measurements determine both the percent polarization
($P\/$) and position angle ($\theta\/$) of the electric field vector
of the linearly polarized light.  For the majority of the program
objects, particularly those with declinations less than 60$^\circ$, we
have multiple observations and can examine whether or not the percent
polarization and/or position angle vary with time.  Our observing runs
were generally a few days in duration and spaced at roughly monthly
intervals.  The ``variability'' column of Table~1 (column 11) contains
a set of codes describing for each object its variability as
determined from our observations. The symbols mean the following:

p $=$ detected significant polarization intrinsic to the source, but
no significant variation in the percent polarization has been
observed.

P $=$ detected significant polarization intrinsic to the source and
the percent polarization has been observed to vary significantly
($\Delta P > 2\sigma_p$).

$\theta_{\rm pref}=$ during the years of monitoring the position
angle of the polarization was not observed to vary significantly
($\Delta \theta < 2\sigma_\theta$).

$\Theta=$ the position angle of the polarization has been confirmed to
be variable ($\Delta \theta > 2\sigma_\theta$).

$\Theta_{\rm pref}$~($\Theta_{\rm stable})~=$ during the years of
monitoring, the position angle of the polarization was observed to
vary significantly, but only over a limited range. We say that it has
a stable polarization position angle if the object's behavior
indicates it might have a preferred position angle, but our
observations do not meet our defined requirements for
designating the object as having a preferred angle of polarization
(see $\S2.2.3 $ for the  definition).

Note that in general the minimum separation between observations is
$\sim$~1~day, so that we are not able to detect or characterize
variations on shorter time scales.  In Figure~1 we present plots of
m$_{\rm v}$, $P$, and $\theta$ versus the date of observation for six
of the monitored objects.

\subsubsection{Maximum Observed Polarization, $P_{\rm m}$}

We will be comparing in paper III the polarization properties of the
XSBLs to various samples of RSBLs, HPQs, and blazars.  One of the
properties we will compare is $P_{\rm m}$, the maximum percent
polarization ever observed for an object.  Ideally we would always
compare polarized fluxes or luminosities, but the advantage of percent
polarization is that it is a quantity that can be measured even under
nonphotometric conditions and does not depend on the availability of a
measured redshift for the object.  When possible we also made
photometric observations which allow the determination of $S_p$, the
polarized flux.  We list in column (3) of Table~1 the maximum observed
white light (unfiltered) percent polarization for each of our program
objects. The variability of the object, thoroughness of monitoring,
and dilution of the polarized nonthermal emission by the unpolarized
host galaxy starlight may affect the observed $P_{\rm m}$.  These
issues will be discussed when we compare the maximum observed
polarizations of XSBLs to those of RSBLs (\S~2.4 and paper III).  When
a limit is listed for $P_{\rm m}$, it is the lowest two sigma limit
(as defined in \S~2.1) obtained for the polarization of the object.

For eleven objects we were able to make polarization measurements
through color filters (see paper I for a description of the filters).
For ten objects the maximum observed polarization at $U$ or $B$ was
greater than the white light $P_{\rm m}$.  In column (5) of Table~1,
we list the maximum observed percent polarization at any optical
frequency if it was greater than the white light measurement.  The
observed values of $P_{\rm m}$ range from 1.3 to 16\%, considerably
below the maximum values observed for some RSBLs and blazars (30 to
40\%).  H~1722+119 has the highest observed percent polarization of
any XSBL ($U$-band polarization of 18\%).

\smallskip

\subsubsection{The Duty Cycle of the Percent Polarization}

If we make the reasonable assumption that all of the XSBLs in the
C-EMSS are intrinsically members of the same class of object, then a
single set of polarization measurements of this sample gives us a
``snapshot'' determination of the duty cycle of polarization for
XSBLs.  We define the duty cycle as the fraction of the sample that
for a given epoch of observation has percent polarizations above a
given cutoff level.  The duty cycle is frequently defined as what
fraction of the time a member of the class is highly polarized (above
the cutoff value). We are making the assumption that the temporal
distribution of polarization in a single object is equivalent to the
distribution of polarization measured in all objects in the class.
Similar calculations have been done for RSBLs (K\"uhr \& Schmidt
1990) and for all blazars (Impey \& Tapia 1990; Fugmann \& Meisenheimer
1988).  Because the majority of our observed two sigma limits for
nondetections are at 4\%, we have chosen that value as the dividing
line in our duty cycle calculation.  This is slightly higher than the
previously adopted boundary of 3\% for detection of significant
polarization (e.g. Moore \& Stockman 1984).  The 3\% value was chosen
as an aid in discriminating against objects polarized by interstellar
dust (typical interstellar values due to dichroic absorption are less
than $< 2$\%; Mathewson \& Ford 1970).  The calculation of the
duty cycles for blazars or RSBLs is not significantly changed if we
use a cut off of 4\%.

For the C-EMSS we have multiple significant observations for twenty of
the twenty-two objects.  We have computed the duty cycle for the
first, second, and last epoch of observation of the C-EMSS sample.
For some objects the first, second, or last epoch of observation is
only an upper limit.  For the duty cycle calculation we have assumed
that these objects were polarized at the limit value.

For objects that were never observed to be polarized (for the C-EMSS
sample, 5), we have considered two possibilities.  We first assume
that these objects are BL Lac objects and that they have just failed
to exhibit the polarized emission when we have observed them.  Under
this assumption we have used the limit values as the observed
polarization.  The second possibility is that these objects are
misclassified and we have also calculated the duty cycle for the
sample without the inclusion of these objects.  For a very few members
of the sample, we do not have three epochs of observation and we have
assumed that the object had a percent polarization greater than 4\%.
For example, we were never able to observe MS~1443.5+6349 and we have
assumed for the calculations below that this object would have had a
polarization greater than 4\%.  We have calculated the duty cycle of
the C-EMSS XSBLs in a manner to avoid at all costs {\it
underestimating} the amount of time spent at large values of percent
polarization. Consequently, we have almost certainly overestimated the
duty cycle.  We calculated the uncertainties in the duty cycle by
considering the one sigma range of observations for each object.  If
an object was observed to be 3.5\% $\pm$0.7\% polarized, the object
was counted as less than four percent, but contributed to the positive
uncertainty.

The computed values for the three epochs are 41$^{+14}_{-27}$\%,
32$^{+14}_{-14}$\%, and 46$^{+10}_{-14}$ for the first, second, and
last observational epoch, respectively.  If we restrict ourselves to
those objects in the C-EMSS which were observed on at least one epoch
to have intrinsic polarized emission, the sample size decreases to
sixteen objects and the new duty cycle estimates are
44$^{+6}_{-31}$\%, 25$^{+19}_{-6}$\%, and 50$^{+13}_{-13}$\%.  The
EMSS is biased against objects of high X-ray flux (see paper I). If we
add the HEAO-A2 objects to the C-EMSS, the duty cycles calculated from
the first and last epochs of observation are 38$^{+11}_{-23}$\% and
42$^{+8}_{-12}$\% (and excluding objects never detected to be
polarized, 38$^{+5}_{-23}$\% and 48$^{+10}_{-10}$\%).  For the entire
program sample (34 objects with detections or two sigma limits of
better than 4\%), the first and last epoch duty cycles are
32$^{+9}_{-9}$\% and 47$^{+9}_{-26}$\%.  If we restrict the
calculation to those objects with confirmed detections on at least one
epoch, our sample size drops to twenty-nine objects and the resulting
duty cycles are 38$^{+10}_{-10}$\% and 45$^{+10}_{-21}$\%.  We adopt
40\%, the average of the values determined for both the EMSS and
C-EMSS plus HEAO-A2 sample, as the duty cycle of XSBLs.

The measured duty cycles are not the consequence of some small
population of highly polarized ($P>4$\%) XSBLs mixed in with a group
of objects that are not capable of being highly polarized.  The
majority of the program objects were observed to be highly polarized
on at least one occasion (20 out of 34).  If we restrict our
consideration to the objects with detected intrinsic polarized
emission (the confirmed BL Lac objects), 69$^{+13}_{-10}$\% were
observed to be highly polarized at least once (20 out of 29, with the
error reflecting that some objects were within one sigma of the
dividing line of 4\% polarization).

\subsubsection{Preferred Polarization Position Angles}

A substantial number of the XSBLs in our monitoring program have
exhibited a preference for a limited range of polarization position
angles.  We describe an object as having a preferred polarization
position angle if during our three years of monitoring the observed
position angles are concentrated to a limited range of angles. This
does not mean that these objects might not lose this preference over
time or develop a new preferred position angle in the decades ahead.
To test for such behavior, even longer periods of monitoring are
required.

While our observing runs were usually separated by a month or more,
many of the objects were observed repeatedly during a two to five day
run.  It could be argued that observations within one week are not
``independent'' and that we should not let four or five observations
in one week skew our impression of the stability of the polarization
position angle.  Unfortunately it is not clear what is enough
separation in time to call two observations ``independent.''  We can,
however, still address the question of preferred position angles over
the time period of our study by limiting ourselves to a consideration
of well-observed objects.  We consider an object to be well-observed
for position angle variability if the following criteria are met:

1.) The object must have been observed to be significantly polarized
during at least six separate observing runs. Observing runs are
generally two to seven days in length and are separated by at least
three weeks.

2.) We must have a significant baseline of observations of the object.
We choose the arbitrary number of 20 months separation between the
first and last observation.

Table~2 lists the XSBLs which meet these criteria.  The objects
MS~0737.9+7441 and H~1722+119 are also included in Table~2 for reasons
explained below.  The individual \hbox{columns} of Table~2 contain the
following information: (1) object name, (2) sample membership, (3)
variability code as defined for Table~1, (4) the number of observation
epochs (nights observed), (5) the number of observing runs during
which the object was significantly polarized, (6) the observed range
of the polarization position angle, (7) the mean position angle
calculated by including only one observation from each observing run,
(8) the variance of the observed distribution of position angles, (9)
the average deviation from the mean, (10) the time in months between
the first and last observations of the object.

For our study, we describe objects which meet the following criteria
as having preferred position angles: 1.) They are well observed for
position angle variability (defined above), 2.) They have average
deviations from the mean angle of less than 20$^\circ$.  This last
criterion is a quantitative means of expressing the observation that
the range of variability of the position angle is limited.  What we
mean by limited range of $\theta$ is best exemplified by the data for
MS~2143.4+0704.  Examination of Figures~1 and 2 shows that over 22
months of observation, the position angle of the polarization of this
BL Lac object did not vary by more than 20$^\circ$.

Under this definition, 11 of the 15 objects in Table~2 have preferred
position angles.  MS~0737.9+7441 and H~1722+119 have observed position
angles which do not vary over a wide range, but do not meet the first
criterion listed above.  We describe them as having stable position
angles.  We only have four epochs of detected polarized emission for
MS~0737.9+7441 and the baseline of observations for both
MS~0737.9+7441 and H~1722+119 is only one year. H~1722+119 was added
late to our study when its identification as an XSBL was available in
1989 (Brissenden \etal 1990).

Of the 22 C-EMSS BL Lac objects, seven are included in our
well-studied sample.  An amazing six out of seven (86\%) have
preferred position angles.  In fact 85\% of the entire well-studied
sample (11 out of 13) have preferred position angles.  All but one of
the objects (MS~0257.9+3429) have confirmed variable position angles,
but the range of variation is limited.  In Figure~2 we present plots
of the normalized Stokes parameters (U/I {\it vs.} Q/I) for the
objects in the well-studied sample.  We have plotted all of our white
light polarization observations for each object.

\subsection{Frequency Dependence of the Polarization}

We were able to obtain multicolor polarimetry for 11 of our
program objects.  A range of frequency-dependent behavior was
observed.  In an effort to quantitatively describe the observed
frequency dependence of the percent polarization (FDP) for these
objects we  use the parameter

$P_\nu = d ( {\rm log} P)/ d ( {\rm log} \nu$).

\noindent{We  characterize the frequency dependence of the polarization
position angle with the analogous parameter}

$\theta_\nu = d \theta/ d ( {\rm log} \nu$).

\noindent{Additional description of these parameters can be found in Smith \&
Sitko (1991).  Table~3 summarizes the frequency-dependent behavior of
the polarization for the observed objects. The column heading symbols
mean the following:}

$P_{\nu}(+)$ = number of observations for which $P_\nu - 2 \times
\sigma(P_\nu) > 0 $. This corresponds to the number of times the
percent polarization was observed to increase (decrease) with the
frequency (wavelength) of observation.

$P_{\nu}(-)$ = number of observations for which $P_\nu + 2 \times
\sigma(P_\nu) < 0 $. This corresponds to the number of times the
percent polarization was observed to decrease (increase) with the
frequency (wavelength) of observation.

$P_{\nu} > 0 =$ number of times $P_\nu > 0$. This is the same
$P_{\nu}(+)$ without the significance restriction.

${\rm max}(P_B/P_I) = $ the maximum ratio between the $B$-band percent
polarization and the $I$-band percent polarization for those
observations where $P_\nu - 2 \times \sigma(P_\nu) > 0 $.

$\vert\theta_\nu\vert = $ number of times that either $\theta_\nu - 2
\times \sigma(\theta_\nu) > 0$ or $\theta_\nu + 2 \times
\sigma(\theta_\nu) < 0$.

${\rm max}(\theta_B - \theta_I) =$ maximum difference (in degrees)
between the $B$-band and $I$-band polarization position angles when
$\theta_\nu - 2 \times \sigma(\theta_\nu) > 0$ or $\theta_\nu + 2
\times \sigma(\theta_\nu) < 0$.

The general results are quite clear.  Whenever significant FDP is
detected the sense is always positive, i.e. the percent polarization
increases with frequency.  Frequency dependence of the polarization
position angle (FD$\theta$) was only detected at better than the 3
$\sigma$ confidence level on one occasion (H~2154$-$304 on 1988 October
29).

\subsection{Dilution of the Polarized Emission by the Host Galaxy}

Most of the XSBLs have evidence of the underlying host galaxy
in their optical spectra as demonstrated by the strength of stellar
absorption features and the 4000\AA~ break in their spectra (Stocke
\etal 1985; Morris \etal 1991). In this subsection we examine the
possible effects of the host galaxy starlight on the observed
polarization of the synchrotron emission from our sample of
BL Lac candidates and objects.

Unpolarized starlight of the elliptical host galaxy (virtually all of
the identified BL Lac object hosts are elliptical galaxies, see for
example Abraham 1991a,b and Ulrich 1989) ``dilutes'' the percent
polarization of the synchrotron source that dominates the observed
properties of BL Lac objects.  Dilution is most pronounced at longer
wavelengths where the intrinsically red spectral energy distribution
of the host galaxy contributes more flux. However, the magnitude of
the effect on the observed polarization will also depend on the
relative brightness and steepness of the power law of the nonthermal
emission.

We have generated model BL Lac object spectra consisting of elliptical
galaxies with an added power law (F$_\nu \propto \nu^{-\alpha}$) and
calculated the observed percent polarization relative to the intrinsic
polarization of the synchrotron source.  Cruz-Gonzalez \& Huchra
(1984) observed BL Lac objects to have a range of power law indices
($\alpha$) of 1.01 to 3.93. The vast majority falling in the range 1.0
to 2.0 and with a mean of 1.89$\pm 0.79$.  H~2154$-$304 has an
observed optical spectral index between 0.7 and 0.8.  We have
generated models with $\alpha$ ranging from the extremely low value of
0.7 to 2.0, close to the mean observed value.  For the elliptical
galaxy we used the spectrum of NGC~4889 provided in Kennicutt (1992)
supplemented at the long and short wavelength ends by the bulge
template of Coleman, Wu, \& Weedman (1980).

The relative brightness of the nonthermal emission to the galactic
starlight must also be set. We have chosen to define this as a
function of the strength of the 4000 \AA~ break in the observed total
flux from the object since this was used as one of the selection
criteria for the EMSS sample of XSBLs (see paper I and Morris \etal
1991).  For our model spectra the ``break strength'' is defined as the
flux difference above and below 4000 \AA~ divided by the flux longward
of 4000 \AA~ (specifically two flux windows were used, 3910 to 3990
\AA~ and 4010 to 4090 \AA). In the EMSS sample all of the objects had
to have 4000 \AA~ breaks less than 25\%.

In Figure 3 we have displayed the ratio of the the observed
polarization to the intrinsic percent polarization of the synchrotron
source for various combinations of power law nonthermal sources and an
elliptical host galaxy for a BL Lac object at a redshift of 0.2.
Three groups of four plots are shown.  Each group corresponds to a
different choice of the strength of the 4000 \AA~ break in the
spectrum of the BL Lac object. Plots are shown for break strengths of
5, 15 and 25\% (the maximum value any object in the EMSS sample could
have). Within each group, the four plots correspond to different
values of the power law spectral index, $\alpha$, of the nonthermal
component.  Note that the various models at a given break strength
(differing spectral index) will have different ratios of nonthermal to
galaxy flux.  The specific values of the spectral index shown in the
figure are $\alpha=$ 0.7, 1.1, 1.5 and 1.9. This choice covers the
bulk of the observed range of spectral indices for BL Lac objects and
includes models that show the {\it maximum} effect that dilution by
the host galaxy can have on the polarization (the $\alpha = 0.7$
models).  We have not included models for large values of spectral
index since in these cases (even for the break strength of 25\%) the
possible effects of dilution become unimportant.  Note that for
smaller break strengths and/or steeper spectral indices (i.e. when the
nonthermal component is a larger fraction of the total emission) the
effects of dilution on the observed polarization properties greatly
decrease.

\subsubsection{How Does Dilution affect $P_{\rm m}$?}

It has been suggested that the low observed percent polarizations of
the XSBLs result from dilution of the polarization by the host galaxy
(Stocke \etal 1985).  This would be consistent with the flux
contribution of the beamed synchrotron component being less for the
X-ray selected objects.  We contend that while the host galaxy is
definitely a greater fraction of the optical emission of the XSBLs,
dilution by starlight is not enough to explain the observed
differences in the polarization properties of X-ray selected and radio
selected BL Lac objects. In this section we discuss arguments that support
the contention that for our sample of XSBLs, the observed distribution
of white light $P_{\rm m}$ is not drastically different from the
distribution of $P_{\rm m}$ for the synchrotron continua free of the
effects of dilution. A major focus of paper III is a quantified
comparison between XSBLs and RSBLs, including the effects of dilution.

For most of the XSBLs with redshifts greater than z$\,=0.2$, $U$ and
$B$-band polarimetry measures the polarization shortward of the 4000
\AA~ in the rest frame of the object.  Assuming that the host galaxy
is an elliptical galaxy, then shortward of the break the object's
emission will be dominated by the nonthermal emission and the observed
percent polarization is a relatively accurate indication of the
intrinsic polarization of the nonthermal source (see Figure 3).  We
have $B$ or $U$ polarimetry of five XSBLs with redshifts greater than
0.2.  In every case, the maximum observed polarization at $U$ and $B$
is within a few percent of the white light value (measured on the same
date).  As we will see in paper III, the observed differences between
the 1 Jy RSBL and C-EMSS XSBL samples are too great to overcome by
shifting the observed \pmax~ distribution of the XSBLs by only a few
percent.

It is also clear from Figure 3 that if a XSBL shows no frequency
dependence in its observed polarization, dilution can not be playing a
major role.  Even for objects at low redshift the observed frequency
dependence is a clue to the amount of dilution that is present (see
$\S 2.4.2$).

Even in the worst possible situation, where the host galaxy is
providing the bulk of the total observed flux and the nonthermal
component has an unrealistically flat spectrum ($\alpha = 0.7$) the
correction factor that needs to be applied to the white light
polarization measurement to get to the intrinsic percent polarization
of the nonthermal source is only a factor of 4 to 5. More typical
correction factors (for breakstrengths of 10 to 15\%, redshifts of
sources 0.1 to 0.4, and power laws of 1.3 to 2.0) are less than a
factor of two.

\subsubsection{Does Dilution Produce the observed FDP?}

The general trends exhibited by the frequency dependence of the
polarization of XSBLs are generally consistent with what is expected
if the optical nonthermal emission from the AGN is diluted by
unpolarized starlight from the host galaxy.  We learned in $\S 2.3$
that if FDP is observed it always is seen with $P_\nu$ positive.  IF
FDP is caused by galactic dilution this would have to be the case (see
Figure 3). Since the galactic starlight has a red spectral energy
distribution the percent polarization is smaller in the red because
the starlight is a larger fraction of the total flux than it is in the
blue.

However, dilution by the host galaxies' starlight can not explain all
of the cases of detected FDP.  H~2154$-$304 is a clear example of an
XSBL for which we know that the host galaxy does not significantly
affect the observed FDP.  Optical and UV polarimetry of this object
has presented very compelling evidence that the FDP observed is
intrinsic to the synchrotron-emitting region (Smith \& Sitko 1991;
Smith et al.  1992; Allen et al. 1993).  Also, any frequency
dependence in $\theta$ can not be explained by starlight of the host
galaxy being included within the observational aperture.  Dilution by
an unpolarized component can only affect the percent polarization, not
the position angle of the polarized component.  Clearly another
mechanism must be responsible for the rare FD$\theta$ observed in
H~2154$-$304 and perhaps in H~0323+022 and H~1219+305.

The optical polarization of the two lowest redshift objects in our
sample, H~0548$-$322 and H~1652+398, are clearly affected by starlight
from their host galaxies.  Given that neither object has shown
significant FD$\theta$, the contribution of the stellar component must
be carefully taken into account before advancing any claims for FDP
intrinsic to the AGN components of these objects.

Another point to keep in mind is that the host galaxy light is
presumably not variable over the short time scales  that we
have been monitoring these objects. Therefore changes in FDP for a
given object must be due to changes in the polarized nonthermal
emission.  For example, if an XSBL has no FDP when it is faint, but
has strong FDP at a later epoch when the object is brighter, then the
observed FDP must be due to the synchrotron source.  In fact, if
dilution were the only mechanism producing FDP there should be a
correlation with the presence of FDP and the apparent magnitude of the
object. Specifically, as the nonthermal source gets fainter, and
assuming that its polarization properties are not strongly dependent
on the luminosity of the source, we might expect that the FDP, as
measured by the ratio of $P_B$/$P_I$, should increase as the host
galaxy becomes a larger fraction of the observed emission.

It is also interesting to note that, assuming galactic dilution is the
only mechanism producing the observe FDP, as $P_\nu$ gets closer to 0
(equivalently, as $P_B/P_I$ gets closer to 1) the affect of dilution
on the determination of the intrinsic polarization of the synchrotron
source goes down. In other words, for objects with little or no
frequency dependence in the observe polarization the white light
polarization measurement must be close to being an accurate
measurement of the intrinsic polarization of the nonthermal light
reaching the observer.

Unfortunately, for most of the objects in our sample we either lack
the necessary data or S/N in the filtered observations to definitively
determine if the observed FDP is caused by galactic dilution and/or
mechanisms intrinsic to the source of nonthermal emission.

\vfil
\eject

\section{Photometry of XSBLs}

\subsection{Are XSBLs Photometric Variables?}

One of the defining criteria of BL Lac objects is that they exhibit
flux variability.  Our photometry of XSBLs allows us to look for
variability and to determine polarized fluxes.  The weather permitted
accurate $V\/$-band photometry on 55 of the 101 nights of
observations.  On some occasions multiband photometry was obtained for
the brighter objects.  Other researchers have also been monitoring the
variability of these objects.  Stocke (1990) reports that all of the
C-EMSS BL Lac objects show variability.  In the ``variability'' column
of Table~1 we indicate with an ``M'' objects for which we have
confirmed variability from our data alone.  We consider that two
photometric observations of an object which differ by more that two
sigma confirm the variability of the object. If an ``M?'' appears it
means that there is a question about the significance of the
variability and the notes on individual objects (paper I) should be
consulted.  Our data, combined with the observations of Stocke \etal
(1991), allow the determination that all of the objects in Table~1 are
photometric variables except for MS~0419.3+1943, H~1101$-$232,
MS~1207.9+3945, H~1426+428, MS~2336.5+0517, MS~2342.7$-$1531, and
MS~2347.4+1924. Continued or improved monitoring might detect
variability in these latter objects since the long term (greater than
two years) behavior of XSBLs is unknown.

We did not observe large or rapid changes in the brightness of these
objects.  When we did detect variability, the change in brightness was
always less than 1.2 magnitundes (peak to peak).  In fact, only
H~1219+305 was observed to vary by more than one magnitude ($\Delta V
= 1.18$).  None of the objects could be considered to be optically
violent variables.  Our photometry is not extensive enough to
determine the typical time scales for photometric variability.

\subsection{Dependence of Polarization on Total Observed Flux}

We have examined our data set to answer the following question: Are
increases in percent polarization accompanied by an increase in total
and/or polarized flux? There is no fixed rule.  Objects are observed
to get brighter, have the polarized flux increase, and have $P$
decrease. Objects are observed to get fainter, have the polarized flux
decrease, and have $P$ increase.  On other occasions, an increase in
unpolarized flux is accompanied by a commensurate increase in
polarized flux, resulting in an increase of brightness, but no change
in the percent polarization.

 While a range of behavior is observed, there is a tendency for the
maximum observed percent polarization to indicate when the object is
experiencing a period of maximum production of polarized flux.
However, this might not be significant since higher percent
polarizations will yield higher F$_p$'s for objects with limited
photometric range.

The data supporting the discussion below are \hbox{presented} in the
following columns of Table~1: (6) the apparent magnitude (m$_{\rm v}$)
of the object when the maximum percent polarization ($P_{\rm m}$) was
observed, (7) the white light percent polarization of the object at
epoch of its brightest emission (m$_{\rm v,Br}$ or m$_{\rm v}$
Brightest), (8) the brightest $V\/$-band magnitude reached by the object
(9) the faintest $V$ magnitude observed for the object (m$_{\rm
v,Fa}$).  When one of these columns has no value it means there are
insufficient data. If a value is given for m$_{\rm v,Br}$ but not for
m$_{\rm v,Fa}$, the value in column (8) is not the brightest
magnitude, but whatever $V\/$-band photometry we had available.  We have
the necessary polarimetry and photometry data to compare the polarized
flux at the epoch of maximum observed percent polarization ($P_{\rm
m}$) and the polarized flux when the object is at its observed
brightest for 18 objects.  We observe $P_{\rm m}$ at the same time as
the brightest emission for nine objects. The epochs of $P_{\rm m}$ and
brightest emission do not correspond for the other objects.  The
polarized flux was greater for two of these objects at the time of
m$_{\rm v,Br}$ despite a lower observed percent polarization ($P$ at
the epoch of m$_{\rm v,Br}< P_{\rm m}$).  For the other seven objects
the polarized flux was larger at the time of $P_{\rm m}$ even though
the corresponding m$_{\rm v}$ is fainter than m$_{\rm v,Br}$.  The
behavior of two specific objects is worth separate attention.
MS~1221.8+2452 has displayed a great range in polarization while
maintaining the same brightness.  On two occasions this object had an
m$_{\rm v}$ of 17.3. The polarizations, however, were quite different,
11.86 $\pm 0.61$~\% and 2.08 $\pm 0.28$~\%.  1E~1415.6+2557 had its
largest output of polarized emission when it was at its faintest. We
do not have a clear picture of what happens to these objects at low
total and polarized fluxes, because it is not easy to obtain adequate
data when these objects are faint and/or weakly polarized.

In general XSBLs are variable in total and polarized emission, but
total flux increases are not necessarily accompanied by an increase in
polarized flux.

\vfill\eject
\section{Are XSBLs Really BL Lac Objects?}

We have obtained a large amount of data on the polarizations and flux
variations of our study sample of XSBLs and can use these data to
examine the objects' classifications as BL Lac objects.  BL Lac
objects must exhibit flux variations and produce variable polarized
emission in addition to the spectroscopic restriction of having no
strong emission lines.  We now reevaluate the classification of each
of the objects.

The usual dividing line for ``significant'' polarization or membership
in the class of BL Lac objects, highly polarized quasars, or blazars
is $P> 3$\%~(Impey \& Tapia 1990; K\"uhr \& Schmidt 1990).  If we
rigorously apply this threshold, several of the objects which were
detected to be polarized would have to be excluded from our list of
XSBLs.  However, the main purpose of the polarization criterion of our
BL Lac object definition is to confirm the synchrotron source
contribution to the flux of the candidate BL Lac object.  Since we
have checked for interstellar polarization and in light of the low
duty cycle of XSBLs (\S~2.2.2), we will retain the objects with low
polarizations unless significant interstellar polarization was
detected.  Note that some researchers have chosen to drop the
requirement that BL Lac candidates be significantly polarized, relying
instead on the overall spectral energy distribution of the sources
(e.g. Schacter \etal 1993).  It remains to be determined if the
objects selected in this manner have the same polarization and
variability properties as the samples studied in this paper and
paper~III.

After applying our definition of a BL Lac object to our study sample,
we are left with 27 confirmed BL Lac objects.  We indicate in column
(12) of Table~1 whether or not the object is a confirmed (C) BL~Lac
object.  Of the 22 C-EMSS BL Lac objects we can only confirm the
classification of 15.  Of the remaining seven, one was not
successfully observed for polarization.  The other six objects failed
to have detectable polarized emission despite repeated observations.
Of the 12 EMSS candidate BL Lac objects, we have detected polarized
emission from three of the six observed objects (MS~2336.5+0517,
MS~2342.7$-$1531, MS~2347.4+1924).

\section{Discussion and Summary of Results}

Our observations have produced the largest and most systematically
compiled database on the polarization of XSBLs.  Our polarimetry
observations confirm that X-ray flux limited surveys find objects that
meet the definition of being BL Lac objects since the majority
(although not all) of the XSBL candidates possess intrinsically
polarized and variable optical continua.  Repeated observations of the
candidate BL Lac objects that we have not extensively monitored should
be continued.  If we had used only the first epoch observations, only
35\% of the sample would have been confirmed to be BL Lac objects.

While these objects are polarized, we have learned that their
polarized emission has a low duty cycle, $\sim$~40\%.  Not only are
they seldom highly polarized, but the maximum observed polarizations
are only around 10\%.  These low values of percent polarization are
not solely the consequence of dilution of the nonthermal component by
the host galaxy starlight.

Some of the objects are observed to have strong frequency dependence
in their percent polarization.  In all such cases the percent
polarization increased with increasing frequency. For many of the
objects in our sample the observed FDP is consistent with a single
synchrotron source in a host elliptical galaxy.  However, there are
examples, most notably H~2154$-$304 where dilution of the polarization
by the host galaxy light can not explain the observed FDP.  Frequency
dependence of the position angle was rarely observed in our sample.
Note that FD$\theta$ can not be explained by a galactic dilution
model.

There is no simple correlation observed between total and polarized
flux, although usually an observed maximum in percent polarization
accompanies the maximum production of polarized flux.

The vast majority (85\%) of the XSBLs have preferred angles for their
optical polarization during the three year monitoring campaign.  As we
have previously discussed (Jannuzi 1990), this must reflect long term
stability of the projected (on the plane of the sky) geometry of the
region producing the polarized emission.  If optical synchrotron
emission is responsible for the production of the polarized emission
(see Kartje \& K\"onigl 1991 for discussion of a accretion disk
scattering model) and this radiation is produced by a jet of material
with a relativistic bulk velocity, then we would expect objects viewed
at significant angles to the relativistic jet to have preferred angles
in contrast to objects viewed directly ``down'' the jet.  We would
also expect a greater incidence of detectable optical jets among the
objects with preferred position angles.  We note that recently only the
second example of an optical jet in a BL Lac object was reported by
Romanishin (1992) (The first being PKS~0521$-$36, Keel 1986; Macchetto
\etal 1991).  The object, 1E~1415.6+2557, is an XSBL and one of the
objects we observe to have a preferred angle for its optical
polarization. The position angle of the jet is $146^\circ$, more
orthogonal than aligned with the mean position angle of the polarized
optical emission ($18^\circ$, see Table 2 and Figure 2).  The
remaining uncertainty on the viewing angle to the jet makes it
difficult to uniquely determine the polarization mechanism. However,
if synchrotron radiation is the mechanism (which is certainly
consistent with the observed variability of the polarized flux), then
the observed preferred angle for the polarization and the observed
``jet'' might be used to constrain models for the magnetic field
distribution in the jet.  If the generation of the optical and radio
``jets'' are closely related and have the same or general geometry we
might also expect a greater incidence of extended and/or asymmetric
radio emission from the objects with preferred angles.

We have confirmed the variability of many of the objects in our
sample.  We have not observed any examples of the large and rapid
fluctuations typical of optically violent variable quasars or the well
known radio selected BL Lac objects. The largest $V\/$-band variation (peak to
peak during the period of monitoring) was only  1.18 magnitudes.

In paper III we compare the observed properties of XSBLs with those of
other extragalactic objects which exhibit significant variable
polarization.  We will discuss further the derivation of the range of
intrinsic physical properties of the class from the observed
properties of BL Lac objects.

\acknowledgments
P.S.S. acknowledges support from NASA grant NAG 5-1630 and NASA
contract NAS 5-29293.  We thank John Stocke, Simon Morris, and the
rest of the EMSS collaboration for making the positions of the EMSS
XSBLs available prior to publication. We acknowledge valuable
discussions with Richard Green and Chris Impey. We thank the
Universities of Minnesota and California, San Diego, for access to the
UCSD/UM 1.52~m and the use of the Two-Holer photopolarimeter.  We thank
Gary Schmidt for building and maintaining Two-Holer, which was
partially funded by NSF grant AST-86-19296.  We thank Margaret Best
for assistance in preparing the tables.

\clearpage
\begin{table}
\begin{tabular}{l}
Table 1 is a landscape table and is contained in a separate file.\\
\end{tabular}
\caption{}
\end{table}
\clearpage

\begin{table}
\begin{tabular}{l}
Table 2 is a landscape table and is contained in a separate file.\\
\end{tabular}
\caption{}
\end{table}
\clearpage

\begin{table}
\begin{tabular}{l}
Table 3 is a landscape table and is contained in a separate file.\\
\end{tabular}
\caption{}
\end{table}

\clearpage
\centerline{\bf FIGURE CAPTIONS}
\bigskip
\noindent
Fig. 1.---Light curves of the apparent $V$-band magnitude
(m$_{\rm v}$), white light (unfiltered) percent polarization ($P$),
and polarization position angle ($\theta$) for six XSBLs. The bottom
axis indicates the UT date of observation.  The top axis is the Julian
Day of the observation minus 2,447,000. The vertical lines centered on
each data point indicate the one sigma uncertainties in the
measurements.  (a.) H~0323+022 and H~1219+305. (b.) MS~1221.8+2452 and
MS~1402.3+0416. (c.)  H~1652+398 and MS~2143.4+0704.

\noindent
Fig. 2.---We have plotted all of our white light polarization
observations for 15 of the objects in our sample.  The normalized
Stokes parameters (Q/I and U/I) are shown with one sigma
uncertainties.  The distance from the origin indicates the percentage
of polarization.  The angle from the x axis is equal to
$2\times\theta$ (where $\theta$ is the PA of the polarized emission).
Note that the vast majority of objects have preferred angles for the
polarization position angle.

\noindent
Fig. 3.---In this figure we display the ratio of the
the observed polarization to the intrinsic percent polarization of the
synchrotron source versus the log of the observed frequency for
various combinations of power law nonthermal sources and an elliptical
host galaxy at a redshift of 0.2.  Three groups of plots are shown for
4000~\AA~ break (rest frame) strengths of 5, 15 and 25\%. Within each
group the four curves correspond to different values of the power law
index, $\alpha$, of the nonthermal component.  The specific values
shown are 0.7, 1.1, 1.5 and 1.9. This choice cover the bulk of the
observed range of $\alpha$ for BL Lac objects.  The unpolarized
starlight of the elliptical host galaxy reduces or dilutes the percent
polarization of the synchrotron source.  The dilution is most
pronounced at longer wavelengths where the intrinsically red spectral
energy distribution of the host galaxy contributes more flux. Note
that as the break strength decreases (i.e.  the nonthermal component
becomes a larger fraction of the total emission) the effects of
dilution greatly decrease. See $\S 2.4$ for further discussion.
The centers of the  filter bands used in the multiband polarimetry are
indicated with the letters $U B V R I$.
\end{document}